\documentclass[letterpaper, 10 pt, conference]{ieeeconf}

\IEEEoverridecommandlockouts
\usepackage{cite}
\usepackage{amsmath,amssymb,amsfonts}
\usepackage{algorithmic}
\usepackage{graphicx}
\usepackage{textcomp}
\usepackage{bbm}
\def\BibTeX{{\rm B\kern-.05em{\sc i\kern-.025em b}\kern-.08em
		T\kern-.1667em\lower.7ex\hbox{E}\kern-.125emX}}

\newtheorem{theorem}{{Theorem}}
\newtheorem{lemma}[theorem]{{Lemma}}

\newtheorem{fact}{{Fact}}
\newtheorem{IEEEproof}{{Proof}}

\title{\LARGE \bf
	Distributed Mechanism Design for Multicast Transmission
}

\author{Nasimeh Heydaribeni and Achilleas Anastasopoulos%
	\thanks{This work was supported in part by NSF Grant ECCS-1608361.}
	\thanks{The authors are with the Department of Electrical Engineering and Computer Science, University of Michigan, Ann Arbor, MI, 48105 USA {\tt\small {heydari,anastas}@umich.edu}}
}
\begin{document}
	
	\maketitle
	\thispagestyle{empty}
	\pagestyle{empty}
	

	\maketitle
	
	\begin{abstract}
		In the standard Mechanism Design framework (Hurwicz-Reiter~\cite{hurwicz2006designing}), there is a central authority that gathers agents' messages and subsequently determines the allocation and tax for each agent.
		We consider a scenario where, due to communication overhead and other constraints, such broadcasting of messages to a central authority cannot take place. Instead, only local message exchange is allowed between agents. As a result, each agent should be able to determine her own allocation and tax based on the messages in the local neighborhood, as defined by a given message graph describing the communication constraints.
		This scenario gives rise to a novel research direction that we call ``Distributed Mechanism Design".
		In this paper, we propose such a distributed mechanism for the problem of rate allocation in a multicast transmission network.
		The proposed mechanism fully implements the optimal allocation in Nash equilibria and its message space dimension is linear with respect to the number of agents in the network.
	\end{abstract}
	
	\begin{keywords}
		mechanism design, rate allocation, decentralized optimization, strategic users, Nash equilibrium
	\end{keywords}
	
	\section{Introduction}
	Most of today's networks consist of a large number of heterogenous agents who have privacy constraints and may act strategically. From the viewpoint of the designer/operator of such networks, solving a resource allocation problem that maximizes the social welfare of the whole network is a difficult task since agents may not be willing to share some of their private information related to their utilities. Appropriate incentives have to be put in place to induce agents to reveal their private information relevant to the welfare optimization problem.
	An appropriate mathematical framework for this setting is mechanism design that has been widely utilized in such areas of research as market allocations~\cite{hurwicz1979outcome,groves1977optimal,yang2005revenue},  rate and resource allocations~\cite{maheswaran2004social,sinha2017mechanism,jain2010efficient,kakhbod2013correction}, data security~\cite{khalili2018designing}, etc.

	In the standard Mechanism Design framework (Hurwicz-Reiter~\cite{hurwicz2006designing}) it is required that agents transmit their messages to a central authority, which in turn,  determines allocation and tax/subsidy for them. Equivalently, it is assumed that agents broadcast their messages to each other and the central authority and everyone can evaluate allocation and taxes for everyone else.
	The motivation for this work is the realistic scenario where such message transmission to a central authority (or equivalently, broadcasting of messages) cannot take place due to network communication constraints. Indeed, in a large network, this may result in a significant communication overhead even for small messages spaces.
	To investigate this problem, we consider a setting in which agents only transmit their messages to their neighboring agents, where neighborhoods are defined through an underlying message graph. Consequently, allocation and tax of each agent must only depend on her neighbors' messages. In other words, each agent can determine her allocation and tax based on the messages she hears and therefore, there is no need for a central authority. This implies that, unlike standard mechanisms, the designed allocation and tax functions cannot have the whole message space as their domain; instead the allocation and tax function for each agent should only depend on the neighborhood messages.
	This additional restriction gives rise to a new research direction that we call ``Distributed Mechanism Design" (DMD).
	A complementary view of DMD stems from the literature of distributed optimization (e.g.,  \cite{nedic2009distributed,boyd2011distributed}). In distributed optimization agents do exchange local messages in order to solve a centralized allocation problem. It is assumed however, that agents are not strategic--infact they are automata--and execute a predefined message exchange algorithm.
	DMD can be thought of as the generalization of distributed optimization to account for strategic agents, i.e. for settings where we can no longer assume that agents will execute a distributed message passing algorithm unless the designer puts in place appropriate incentives for them to do so.
	Note that DMD is not to be confused with the literature of distributed optimization that attempts to resolve ``privacy'' issues by means of dithering (i.e. adding noise to) the exchanged messages as in~\cite{HuMiVa15}.
	In that area of research agents are given some privacy guarantees, but are still considered non-strategic automata.

	In this paper, we propose a distributed mechanism for rate allocation in a multicast transmission network. A non-distributed mechanism for efficient allocation in multicast networks has been proposed in~\cite{kakhbod2013correction,sinha2017mechanism} and our model closely follows these works.
	The current work builds on a distributed mechanism for Walrasian and Lindahl allocation in private and public goods, respectively,  that was proposed in~\cite{sinha2017distributed}\cite[Ch. 4]{sinha2017absi}, as well as the distributed mechanism for unicast networks proposed in~\cite{HeAn18}.
	The contributions of this paper are as follows. The proposed mechanism is (a) distributed, it (b) fully implements the optimal allocation in Nash equilibria (NE) (i.e. there are no extraneous equilibria), and (c) its message space dimension is linear with respect to the number of agents in the network.
	Furthermore, the mechanism is (d) individually rational and (e) weak budget balanced at the NE.
	We have utilized an idea similar to the radial allocation~\cite{yang2005revenue,maheswaran2004social,sinha2017mechanism} to achieve feasibility at NE. Unlike the mechanism in~\cite{sinha2017distributed} that defines messages with dimensionality per user growing linearly with the number of users, in this work, the message dimensionality of each agent is linear with respect to the size of her neighborhood and this is a result of utilizing ``summary" messages (see~\cite[Ch. 4]{sinha2017absi},\cite{HeAn18}).

	The rest of the paper is structured as follows. In Section~\ref{section2}, the model and problem formulation are discussed. Section~\ref{section3} presents the proposed distributed mechanism by stating all of the properties of message communication network and defining allocation and tax functions. In Section~\ref{section4}, the properties of the designed mechanism are explained  and the main results are presented. In Section~\ref{sectionV}, an alternative mechanism is discussed for relaxing an assumption on the message network. We conclude in Section~\ref{section6} with some comments on message dimensions and a discussion of the results.
	The proofs of the intermediate lemmas can be found in the extended version of this paper~\cite{ARXIV-VERSION}.

	\section{Model} \label{section2}
	
\begin{figure}[h]
		\centering
		\includegraphics[width=7.3cm]{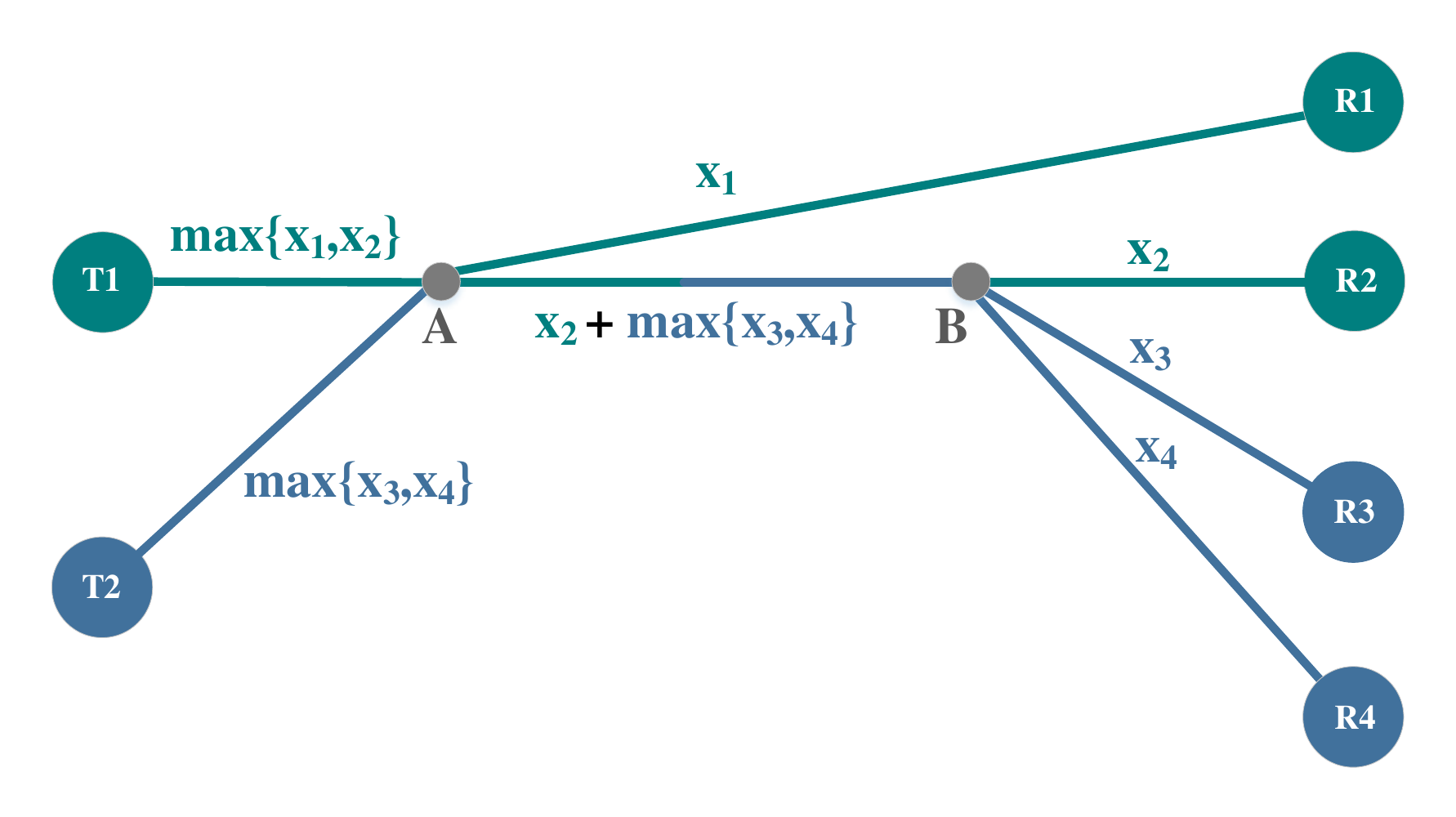}
		\vspace{-0.3cm}
		\caption{Network with multicast transmission. Even though both R1 and R2 use link T1-A, it is only loaded with the rate $\max\{x_1,x_2\}$ due to the multicast transmission.}
		\label{multicast}
	\end{figure}

	We are following closely the model developed in \cite{sinha2017mechanism}. We consider a multicast network consisting of multiple sources $\mathcal{K}=\{1,...,K\}$ and strategic receivers (which are called agents) $\mathcal{N}=\{1,...,N\}$ in which data is transmitted from each source to multiple agents. Hence, the agents are classified into $K$ groups denoted by $\mathcal{K}=\{1,...,K\}$ based on their data source. The set of agents in group $k$ (agents having the same source $k$)  are denoted by $\mathcal{G}_k$ and $G_k$ is the number of them. The group of agent $i$ is denoted by $k(i)$. Since agents in each group receive data from the same source, one common data stream can be transmitted in each link shared by some of  the agents of the same group and the rate of the common data stream is the maximum of the demanded rates by those agents. In other words, every source transmits the common data of agents of its group in each link by the best quality demanded and each agent can regenerate her own data by sampling from the received data stream to get her desirable quality. This scenario is as if agents inside a group share the bandwidth with each other but they have competition for it with other groups which is referred to as intergroup competition and intragroup sharing in \cite{sinha2017mechanism}. One example of multicast transmission is illustrated in Fig.~\ref{multicast}.
	
	The network links are denoted by $\mathcal{L}=\{1,...,L\}$, each of which has capacity $c^l$. Agent $i$'s data stream is transmitted via links $\mathcal{L}_i \subset \mathcal{L}$ with $|\mathcal{L}_i|=L_i$. For each link $l$, agents using it are denoted by $\mathcal{N}^l$ with $|\mathcal{N}^l|=N^l$ and we denote $\mathcal{N}^l \cap \mathcal{G}_k$ by $\mathcal{G}_k^l$. Further, $\mathcal{K}^l$ is the subset of groups that are using link $l$ and its cardinality is denoted by $|\mathcal{K}^l|=K^l$. We assume $K^l\geq 2$ that is at least two groups use each link $l$ and this is for having competition in using each link.

	The data rate of agent $i$ is denoted by $x_i$ and the vector of allocated rates is $x=(x_1,...,x_N)$. Agent $i$ has a valuation over her data rate that is modeled by the function $v_i(x_i)$. The following assumptions are imposed on the valuation functions. We know that for every $i \in \mathcal{N}$, $v_i(.) \in \mathcal{V}_0$, where $\mathcal{V}_0$ is the set of strictly concave, monotonically increasing, twice differentiable and $\mathbb{R_+}\rightarrow \mathbb{R}$ functions with continuous second derivatives.
	
	The designer's goal is to maximize the social welfare, which is the summation of the agents' valuations, by determining the efficient $x$ that is consistent with the capacity constraints of network and the multicast aspect. Therefore, we formulate the following optimization problem,
	\begin{subequations}\label{CP}
		\begin{align}
			&\max_x \sum_{i \in \mathcal{N}} v_i(x_i)  \label{CP-a}\\
			\textrm{s.t.} \quad \ & x_i \geq 0 \quad \forall i \in \mathcal{N} \label{CP-b}\\
			\textrm{and} \quad \  & \sum_{k\in \mathcal{K}^l} \max_{i \in \mathcal{G}_k^l} \{x_i\} \leq c^l \quad \forall l \in \mathcal{L}  \label{CP-c}.
		\end{align}
	\end{subequations}
	
	
	Generally, problem \eqref{CP} can also model unicast transmission and a number of other scenarios of network utility maximization with linear inequality constraints.
	\subsection{Necessary and Sufficient Optimality Conditions}
	We utilize KKT conditions to characterize the solution of problem \eqref{CP}, but first we need to rewrite it in another form to change it to a convex optimization problem. We introduce the variable $b_k^l$ for each $l \in \mathcal{L}$ and $k \in \mathcal{K}^l$ that represents the maximum demand of agents in group $k$ that use link $l$. It is straightforward to show that problem \eqref{CP} and the one below are equivalent,
	\begin{subequations}\label{CP2}
		\begin{align}
			&\max_x \sum_{i \in \mathcal{N}} v_i(x_i)  \label{CP2-a}\\
			\textrm{s.t.} \quad \ & x_i \geq 0 \quad \forall i \in \mathcal{N} \label{CP2-b}\\
			\textrm{and} \quad \  & \sum_{k\in \mathcal{K}^l} b_k^l \leq c^l \quad \forall l \in \mathcal{L}  \label{CP2-c}
			\\
			\textrm{and} \quad \  & x_i\leq b_k^l \quad \forall  l \in \mathcal{L} , \ k \in \mathcal{K}^l, \ i \in \mathcal{G}_k^l. \label{CP2-d}
		\end{align}
	\end{subequations}
	In this problem, the valuation functions are concave and all of the constraints are affine. Therefore,  problem \eqref{CP2} is a convex optimization problem and hence, KKT conditions are necessary and sufficient for its solution. We use dual variables $\lambda$ and $\mu$ as follows.  $\lambda=\{\lambda_l, l\in \mathcal{L}\}$, each of which corresponds to one of the constraints in \eqref{CP2-c} and $\mu=\{\mu_i^l, \forall l \in \mathcal{L}, i \in \mathcal{N}^l\}$, each of which corresponds to one constraint in \eqref{CP2-d}. We can write KKT conditions at optimal point $(x^*, b^*, \lambda^*, \mu^*)$ as
	\renewcommand{\labelenumi}{(\alph{enumi})}
	\begin{enumerate}
		\item Primal Feasibility: $x^*$ and $b^*$ satisfy  \eqref{CP2-b} and  \eqref{CP2-c} and \eqref{CP2-d}.
		\item Dual Feasibility: $\lambda_l^*\geq 0 \ \forall l\in\mathcal{L}$ and ${\mu_i^l}^* \geq 0 \  \forall l \in \mathcal{L}, i \in \mathcal{N}^l$.
		\item Complimentary Slackness:
		\begin{subequations} \label{eq:kkt}
			\begin{equation}
				\lambda_l^*(c^l -\sum_{k\in \mathcal{K}^l} {b_k^l}^*)=0 \quad \forall l\in\mathcal{L},
			\end{equation}
			\begin{equation}
				{\mu_i^l}^*(x^*_i-{b_k^l}^*)=0 \quad \forall l \in \mathcal{L}, k \in \mathcal{K} ^l, i \in \mathcal{G}_k^l.
			\end{equation}
			\item Stationarity:
			\begin{equation}
				v_i^\prime(x_i^*)=\sum_{l\in\mathcal{L}_i}{\mu_i^l}^* \quad \forall i \in \mathcal{N} \quad \textrm{if} \quad x_i^*>0,
			\end{equation}
			\begin{equation}
				v_i^\prime(x_i^*)\leq\sum_{l\in\mathcal{L}_i}{\mu_i^l}^* \quad \forall i \in \mathcal{N} \quad \textrm{if} \quad x_i^*= 0,
			\end{equation}
			\begin{equation}
				\lambda_l^*=\sum_{i \in \mathcal{G}_k^l} {\mu_i^l}^* \quad \forall  l \in \mathcal{L}, k \in \mathcal{K}^l.
			\end{equation}
		\end{subequations}
	\end{enumerate}

	\section{Distributed Mechanism} \label{section3}
	
	The designed mechanism consists of a message space $\mathcal{M}_i$ for every agent  $i \in \mathcal{N}$ and allocation and tax functions that are denoted by $\hat{x}_i(.)$ and $\hat{t}_i(.)$, respectively. These functions depend only on neighboring agents' messages and this is why this mechanism is ``distributed". We can characterize the mechanism completely by specifying the tuple  $(\mathcal{M},(\hat{x}_i(.))_{i \in \mathcal{N}},(\hat{t}_i(.))_{i \in \mathcal{N}})$ where $\mathcal{M}=(\mathcal{M}_1\times...\times\mathcal{M}_N)$. This mechanism induces a game $\mathfrak{G}=(\mathcal{N},\mathcal{M},(\hat{u}_i(.))_{i \in \mathcal{N}})$, where the utility functions are $\hat{u}_i(m)=v_i(\hat{x}_i(m))-\hat{t}_i(m)$. The set of all Nash equilibria of the game $\mathfrak{G}$ is denoted by $\mathcal{NE}$.
	
	\subsection{Message Network}
\begin{figure}[h]
		\centering
		\includegraphics[width=7.3cm]{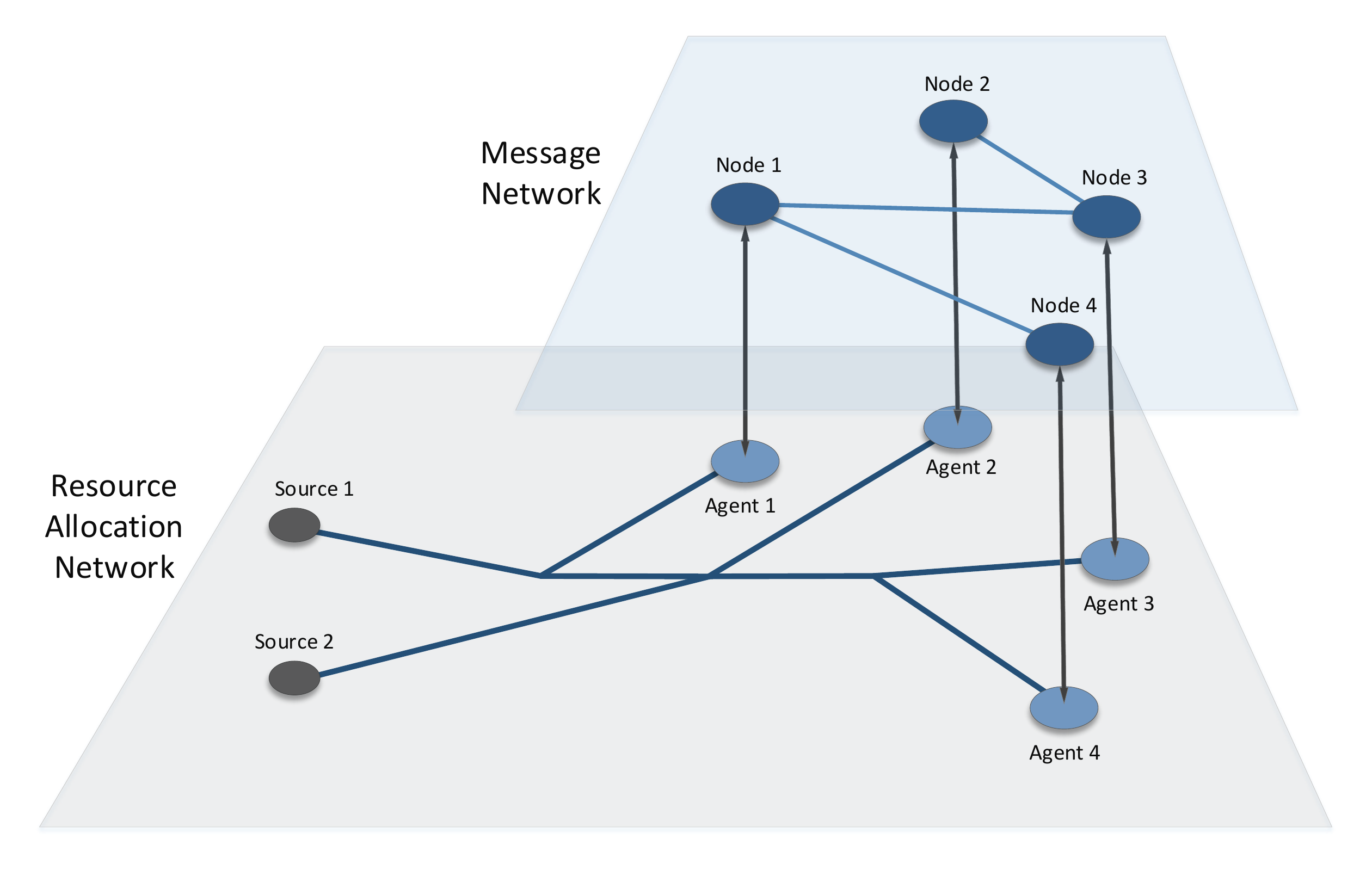}
		\caption{Message network vs. resource allocation network}
		\label{twolayer}
	\end{figure}

	As mentioned earlier, the mechanism is distributed in the sense that message transmission is done locally. This is modeled by a message transmission network that is an undirected graph in which agents are denoted by nodes and an edge between two agents indicates that these two agents hear each others' messages. Otherwise, they do not have access to each others' messages due to communication (or complexity) constraints. This network is called ``message network". Notice that the message network is different from the data transmission network related to problem \eqref{CP}. The message network enables the decentralized solution of that problem and is relevant even for the more general scenarios that can be modeled by problem \eqref{CP}. In Fig.~\ref{twolayer}, the two networks are illustrated.
	
	We consider an arbitrary spanning tree on the  graph of the message network and assume message transmission is done via this tree which will be referred to as the ``message graph" and denoted by $\mathcal{GR=(N,E)}$. For all $i\in \mathcal{N}$, $\mathcal{N}(i)$ is the set of neighbors of agent $i$ in $\mathcal{GR}$ and $|\mathcal{N}(i)|=N(i)$. Further, $n(i,j)$ is agent $i$'s neighbor which is on the shortest path from $i$ to $j$. Also, $\mathcal{N}^l(i)$ denotes the set of agents in $\mathcal{N}(i)$ using link $l \in \mathcal{L}_i$ and $|\mathcal{N}^l(i)|=N^l(i)$. For each agent $i \in \mathcal{N}$, the function $\Phi(i)$ arbitrarily chooses one agent $j \in \mathcal{N}(i)$ and we define the set $I_i=\{h\in \mathcal{N}(i): \Phi(h)=i\}$.  The role of this function will become evident in the rest of this section where we describe the allocation and tax functions.

	The following assumption is imposed on the message graph for simplicity of exposition.
	\newtheorem{assumption}{Assumption}
	\begin{assumption}
		For each link $l \in \mathcal{L}$, the sub-graph consisting of agents $i \in   \mathcal{N}^l$ is a connected graph. Also, for each link $l \in \mathcal{L}$ and group $k \in \mathcal{K}^l$, there is at least one node $i \in \mathcal{G}_k^l$  that is connected to all other nodes $j \in \mathcal{G}_k^l$ and is denoted by $c(k,l)$.\label{assump}
	\end{assumption}
	For each agent $i \in \mathcal{N}$, set $\mathcal{C}_i$ is defined as the set of links $l$ for which $c(k(i),l)=i$.
	
	In Section \ref{sectionV}, a relaxed assumption is imposed on the message graph and an alternative mechanism is proposed.

	\subsection{Message Components}
	
	The message $m_i\hspace{-0.1cm}=\hspace{-0.1cm}(y_i,\underline{y_i},n_i,q_i,p_i,w_i,z_i,a_i)$ is quoted by agent $i$ in this mechanism.
	The reason for such a complex message structure stems form the fact that (a) all agents within a group $\mathcal{G}_k^l$ need access to the maximum demanded rate in that group and (b) this information needs to be disseminated to all agents in the network while satisfying the communication constraints.
	In the following, we give intuitive explanations for the meaning of each of the eight message components.
	The first message, $y_i \in \mathbb{R}_+$, is the agent's demanded rate. 
	The second message is defined as $\underline{y_i}=(\underline{y_i}^l, l \in \mathcal{L}_i)\in \mathbb{R}_+^{L_i}$, where, each of the messages $\underline{y_i}^l$ is capturing whether the specific agent belongs in the group of agents that demand the maximum rate within the group $\mathcal{G}^l_{k(i)}$.
	We call these messages as proxies of the ``group demand''. Specifically, at NE, this message will become zero if the agent is not in the max group, and otherwise, it will be equal to the maximum demanded rate of group $\mathcal{G}^l_{k(i)}$ divided by the number of users in the max group.
	The third message, $n_i=(n_i^{j,l}, j\in\mathcal{N}(i) , l\in \mathcal{L}) \in \mathbb{R}_+^{L\times N(i)}$, consists of components $n_i^{j,l}$, each of which is a proxy for the sum of group demands of the agents $h \in \mathcal{N}^l$  with $n(i,h)=j$. These messages are referred to as ``summary'' messages.
	The fourth message,  $q_i=(q_i^j, j \in I_i)\in \mathbb{R}_+^{|I_i|}$, consists of elements $q_i^j$, each of which is a proxy for agent $j$'s demand.
	The fifth message consists of two components, $p_i=({^{1}p}_i,{^{2}p}_i)$. The first component is defined as ${^{1}p}_i=({^{1}p}_i^l, l \in \mathcal{L}_i) \in \mathbb{R_+}^{L_i}$, where each message ${^{1}p}_i^l$ is the price that agent $i$ is willing to pay for using link $l$.
	This is essentially a proxy for the dual variable ${\mu^l_i}^*$ that appears in the KKT conditions~\eqref{eq:kkt}.
	The second component is defined as ${^{2}p}_i=({^{2}p}_i^{j,l}, j\in I_i, l \in \mathcal{L}_j) \in \mathbb{R}_+^{(\sum_{j\in I_i}L_j)}$, where each variable ${^{2}p}_i^{j,l}$ is the price that agent $i$ thinks agent $j$ should pay for using link $l$.
	The sixth message, $w_i=(w_i^l, l\in \mathcal{L}_i) \in \mathbb{R}_+^{L_i}$, consists of components $w_i^l$, each of which is a proxy for the price that group $k(i)$ is willing to pay for link $l$. These messages have to converge at NE to the dual variable ${\lambda_l}^*$ in the KKT conditions~\eqref{eq:kkt} for all users $i\in \mathcal{N}^l$.
	The seventh message is defined as $z_i=({^{1}z}_i,{^{2}z}_i)$. The first component ${^{1}z}_i=({^{1}z}_i^l, \ l \in \mathcal{C}_i)\in \mathbb{R_+}^{|\mathcal{C}_i|}$ consists of elements ${^{1}z}_i^l$, each of which is a proxy for maximum value of demands of agents in $\mathcal{G}_{k(i)}^l$.
	Further, ${^{2}z}_i=({^{2}z}_i^l, \ l \in \mathcal{C}_i)\in \mathbb{R_+}^{|\mathcal{C}_i|}$ consists of elements ${^{2}z}_i^l$, each of which is a proxy for the number of agents that have maximum demand in $\mathcal{G}_{k(i)}^l$.
	Finally, the eighth message, $a_i=({^1a}_i,{^2a}_i)$, consists of two components. The first component, ${^1a}_i=( {^1a}_i^l, l \in \mathcal{L}_i)\in \mathbb{R}_{++}^{L_i}$, is a vector of messages that have technical roles and will be useful in having efficient NE. The second component, ${^2a}_i=( {^2a}_i^{j,l},  j\in I_i, l \in \mathcal{L}_j) \in \mathbb{R}_{++}^{(\sum_{j\in I_i}L_j)}$, consists of the elements ${^2a}_i^{j,l}$, each of which is a proxy for the message ${^1a}_j^l$.

	For each agent $i \in \mathcal{N}$ and  every link $l \in \mathcal{L}$, we define $y_i^l$ as the extension of $\underline{y_i}^l$ to every link $l \in \mathcal{L}$,
	\begin{equation}
		y_i^l=\left\{
		\begin{array}{cc}
			\underline{y_i}^l & \text{if} \ l\in \mathcal{L}_i\\
			0 & \textrm{oth.}
		\end{array}
		\right.
	\end{equation}
	For each agent $i \in \mathcal{N}$ and $l \in \mathcal{L}_i$,  ${^{1}\bar{z}}_i^l$ and ${^{2}\bar{z}}_i^l$ are defined as
	\begin{subequations}
		\begin{align}
			&\hspace{-0.2cm}{^{1}\bar{z}}_i^l=\left\{
			\begin{array}{cc}
				\hspace{-0.2cm}\max\{q_{\Phi(i)}^i,\max_{j \in \mathcal{G}_{k(i)}^l,j \neq i}\{{y}_j\}\} &  \textrm{if} \ l \in \mathcal{C}_i \\
				\hspace{-0.2cm}{^{1}z}_{c(k(i),l)}^l & \textrm{if} \ l \notin \mathcal{C}_i
			\end{array}\right.\\
			&\hspace{-0.2cm}{^{2}\bar{z}}_i^l\hspace{-0.1cm}=\hspace{-0.1cm}\left\{
			\begin{array}{cc}
				\hspace{-0.25cm}\textbf{1}_{\{q_{\Phi(i)}^i\}}({^{1}\bar{z}}_i^l)\hspace{-0.05cm}+\hspace{-0.05cm}\sum_{j \in \mathcal{G}_{k(i)}^l,j \neq i}\textbf{1}_{\{y_j\}}\hspace{-0.05cm}(^{1}\bar{z}_i^l) & \ \hspace{-.4cm}\textrm{if} \ l \in \mathcal{C}_i \\
				\hspace{-0.25cm}{^{2}z}_{c(k(i),l)}^l & \ \hspace{-.4cm}\textrm{if} \ l \notin \mathcal{C}_i
			\end{array}\right.
		\end{align}
	\end{subequations}
	where $\textbf{1}$ is the indicator function ($\textbf{1}_{\{\alpha\}}(\beta)$ is equal to $1$ only if $\alpha=\beta$). We have defined ${^{1}\bar{z}}_i^l$ so that each agent $i \in \mathcal{N}^l$ gets aware of the maximum demand of agents $j \in \mathcal{G}_{k(i)}^l$ at NE, and it is calculated by her as a function of her neighbors' messages. Similarly, at NE, ${^{2}\bar{z}}_i^l$ is the number of agents with maximum demand among all of the agents $j \in \mathcal{G}_{k(i)}^l$.

	\subsection{Allocation Functions}
	We utilize an idea similar to the radial allocation \cite{sinha2017mechanism} to have feasible allocation at NE. With this goal in mind, the allocation function is defined as
	\begin{subequations}
		\begin{equation}
			\hat{x}_i(m)=r_i \  {y_i},
		\end{equation}
		where $r_i$ is agent $i$'s radial allocation factor, $r_i=\min_{l\in\mathcal{L}} \frac{c^l}{f_i^l}$,
		and for $l \in \mathcal{L}_i$, $f_i^l$ is defined as
		\begin{equation}
			f_i^l=
			\frac{q_{\Phi(i)}^i\textbf{1}_{\{q_{\Phi(i)}^i\}}({^{1}\bar{z}}_i^l)}{{^{2}\bar{z}}_i^l}+\sum_{j\in\mathcal{N}(i)}(y_j^l+\sum_{h\in\mathcal{N}(j) , h \neq i}n_j^{h,l}),
		\end{equation}
		and for $l \notin \mathcal{L}_i$, $f_i^l$ is defined as
		\begin{equation}
			f_i^l=\sum_{j\in\mathcal{N}(i)}(y_j^l+\sum_{h\in\mathcal{N}(j) , h \neq i}n_j^{h,l}) .
		\end{equation}
	\end{subequations}
	Since agents don't have access to all of the messages, for each agent $i$, we should define a proxy, $f_i^l$, for the sum of group demands of agents on link $l$  to enable feasible allocation at NE.
	Note that the quantity $f^l_i$ does not depend on agent $i$'s messages.

	\subsection{Tax Functions}
	The tax functions are $\hat{t}_i(m)=\hat{t}_i^{\mathfrak{c}}(m)+\sum_{l\in\mathcal{L}}\hat{t}_i^l(m)$, where 
	\begin{subequations}	
		\begin{equation}
			\hat{t}_i^{\mathfrak{c}}(m)\hspace{-0.1cm}=\hspace{-0.2cm}\sum_{j \in I_i}\hspace{-0.05cm}\sum_{l \in \mathcal{L}_j}\hspace{-0.1cm}(({^{2}p}_i^{j,l}\hspace{-0.05cm}-{^{1}p}_j^l)^2+({^{2}a}_i^{j,l}\hspace{-0.05cm}-{^{1}a}_j^l)^2)\hspace{-0.05cm}+\hspace{-0.15cm}\sum_{j\in I_i}\hspace{-0.05cm}(q_i^j-y_j)^2
		\end{equation}
		and for each component $\hat{t}_i^l(m)$ we have three cases.
		
		For $l \in \mathcal{L}_i , l \notin \mathcal{C}_i$, it is defined as
		\begin{equation}
			\begin{aligned}\label{eq:t1} &\hat{t}_i^l(m)={^{2}p}_{\Phi(i)}^{i,l}\hat{x}_i(m)+\sum_{j\in\mathcal{N}(i)}(n_i^{j,l}-y_j^l-\hspace{-0.4cm}\sum_{h\in\mathcal{N}(j) , h \neq i}n_j^{h,l})^2\\
				&+(\underline{y_i}^l\hspace{-0.1cm}-\hspace{-0.05cm}\frac{q_{\Phi(i)}^i\textbf{1}_{\{q_{\Phi(i)}^i\}}\hspace{-0.05cm}({^{1}\bar{z}}_i^l)}{{^{2}\bar{z}}_i^l})^2
				+\bar{w}_{-i}^l(\hat{w}_i^l-\bar{w}_{-i}^l)(c^l-r_if_i^l)^2\\&+(\hat{w}_i^l-\bar{w}_{-i}^l)^2+{^{2}p}_{\Phi(i)}^{i,l}(^{1}p_i^l-\hspace{-0.05cm}{^{2}p}_{\Phi(i)}^{i,l})({^{1}\bar{z}}_i^l-q_{\Phi(i)}^i)^2
				\\&+(w_i^l-w_{c(k(i),l)}^l)^2.
			\end{aligned}
		\end{equation}
		For $l \in \mathcal{L}_i , l \in \mathcal{C}_i$, we have
		\begin{equation}
			\begin{aligned}
				&\hat{t}_i^l(m)={^{2}p}_{\Phi(i)}^{i,l}\hat{x}_i(m)+\sum_{j\in\mathcal{N}(i)}(n_i^{j,l}-y_j^l-\hspace{-0.4cm}\sum_{h\in\mathcal{N}(j) , h \neq i}n_j^{h,l})^2\\
				&+(\underline{y_i}^l-\frac{q_{\Phi(i)}^i\textbf{1}_{\{q_{\Phi(i)}^i\}}({^{1}\bar{z}}_i^l)}{{^{2}\bar{z}}_i^l})^2+(^{1}z_i^l-{^{1}\bar{z}}_i^l)^2+(^{2}z_i^l-{^{2}\bar{z}}_i^l)^2\\&
				+ (w_i^l-{^{2}p_{\Phi(i)}^{i,l}}-\hspace{-0.4cm}\sum_{j \in \mathcal{G}_{k(i)}^l, j \neq i}\hspace{-0.25cm}{^{1}p_j^l})^2\hspace{-0.05cm}+\bar{w}_{-i}^l(\hat{w}_i^l-\bar{w}_{-i}^l)(c^l-r_if_i^l)^2\\
				&+(\hat{w}_i^l-\bar{w}_{-i}^l)^2\hspace{-0.05cm}+{^{2}p}_{\Phi(i)}^{i,l}(^{1}p_i^l-{^{2}p}_{\Phi(i)}^{i,l})({^{1}\bar{z}}_i^l-q_{\Phi(i)}^i)^2,
			\end{aligned}
		\end{equation}
		where for each link $l$ and agent $i \in \mathcal{N}^l$, $\hat{w}_i^l$ is defined as
		\begin{equation}
			\hat{w}_i^l=\left\{
			\begin{array}{cc}
				\hspace{-0.2cm}\sum_{j \in \mathcal{G}_{k(i)}^l}{^{1}p_j^l} + ({^1a}_i^l-{^2a}_{\Phi(i)}^{i,l}) & \textrm{if} \ l \in \mathcal{C}_i\\[0.2cm]
				\hspace{-0.2cm}w_{c(k(i),l)}^l-{^{2}p}_{\Phi(i)}^{i,l}+{^{1}p}_i^l + ({^1a}_i^l-{^2a}_{\Phi(i)}^{i,l}) & \textrm{if} \   l \notin \mathcal{C}_i
			\end{array}
			\right.
		\end{equation}
		Further, $\bar{w}_{-i}^l$ is defined as
		\begin{equation}
			\bar{w}_{-i}^l=
			\frac{1}{N^l(i)}\sum_{j \in \mathcal{N}^l(i)}w_j^l.
			\label{w-i}
		\end{equation}
		Finally, for $l \notin \mathcal{L}_i$, tax is
		\begin{equation}
			\begin{aligned}
				\hat{t}_i^l(m)=\sum_{j\in\mathcal{N}(i)}(n_i^{j,l}-y_j^l-\sum_{h\in\mathcal{N}(j) , h \neq i}n_j^{h,l})^2.
			\end{aligned}
		\end{equation}
	\end{subequations}
	
	Intuitively, the tax functions provide some penalties to incentivize agents for quoting messages in a desirable manner. With this goal in mind, taxes contain three types of terms. The first type is a rate times price component (e.g., the first term in~\eqref{eq:t1}).
	The second type consists of quadratic terms that at NE will become zero and thus can be thought of as incentivizing agents to come to a consensus (e.g., all terms in~\eqref{eq:t1} other than the first, the fourth and the sixth). This enables the mechanism to provide proxies for the missing information of agents at NE, in addition to the requirements of having efficient allocation at NE. The third type relates to the complimentary slackness conditions in~\eqref{eq:kkt} (e.g., the fourth and the sixth term in~\eqref{eq:t1}).
	The reason of defining different tax functions is that different incentives are required for different agents. For instance, for each link $l$ and $k \in \mathcal{K}^l$, agent $c(k,l)$ should announce the proper messages that convey information about other agents in $\mathcal{G}_k^l$ so that all of them can have consensus on their group demands on link $l$ at NE. Furthermore, each agent $i$ has to pay a tax even for links $l \notin \mathcal{L}_i$, which is required for consensus about the ``summary" messages. The intuition about each tax term will become more evident from the results of Section~\ref{section4}.

	\section{Mechanism Properties}  \label{section4}
	\begin{fact}\label{dist}
		The mechanism  $(\mathcal{M},\hat{x},\hat{t})$ is distributed.
	\end{fact}
	
	This can be observed from the definition of allocation and tax functions that are only generated based on each agent's own messages and her neighboring agents' messages.
	
	\begin{theorem}(Full Implementation, Individual Rationality and Weak Budget Balance)
		At each Nash equilibrium $m \in \mathcal{NE}$ of the game $\mathfrak{G}$, the allocation vector $\hat{x}(m)$ is efficient; i.e. it is equal to the solution, $x^*$, of problem \eqref{CP}. In addition, for each agent, individual rationality is satisfied at all NE. Further, the game $\mathfrak{G}$ is weak budget balanced at all NE.
		\label{FI}
	\end{theorem}
	
	Since $x^*$ is unique and according to Theorem \ref{FI}, for all $m \in \mathcal{NE}$, the allocation vector  $\hat{x}(m)$ is unique.

	Before proving Theorem \ref{FI}, some lemmas are presented that are necessary for its proof. 
	
	\begin{lemma}(Concavity)\label{con}
		The function $\hat{u}_i(m_i,m_{-i})$ is strictly concave w.r.t. $m_i$.
	\end{lemma}
	The strict concavity of $\hat{u}_i(m_i,m_{-i})$ w.r.t. $m_i$ helps us calculate the best response functions by setting the gradient of  $\hat{u}_i(m_i,m_{-i})$ w.r.t. $m_i$  to be equal to zero. Yet, it is not always possible for all of the elements of the gradient vector to be set to zero. In this case, those elements are either always positive or always negative. If any of the elements were always positive, then as message spaces are unbounded from above, there is no best response. Otherwise, if any of the elements of the gradient vector were always negative, the best response would be zero for that element.
	
	\begin{lemma}\label{quadratic}
		At any $m \in \mathcal{NE}$, the following equations hold for any $i \in \mathcal{N}$.
		\begin{subequations}
			\begin{align}
				&q_i^j=y_j, \ \forall j \in I_i\\
				&\underline{y_i}^l=
				\frac{q_{\Phi(i)}^i\textbf{1}_{\{q_{\Phi(i)}^i\}}({^{1}\bar{z}}_i^l)}{{^{2}\bar{z}}_i^l}, \ \forall l \in \mathcal{L}_i\\
				&^{2}p_i^{j,l}={^{1}p}_j^l , \ \forall j \in I_i, l \in \mathcal{L}_j \\
				&w_i^l=\left\{
				\begin{array}{cc}
					\hspace{-0.2cm}w_{c(k(i),l)}^l & \ \hspace{-0.2cm} \textrm{if} \ \ l \in \mathcal{L}_i , l \notin \mathcal{C}_i\\
					\hspace{-0.2cm}{^{2}p_{\Phi(i)}^{i,l}}+\sum_{j \in \mathcal{G}_{k(i)}^l, j \neq i}{^{1}p_j^l} & \ \hspace{-0.2cm} \textrm{if} \ \  l \in \mathcal{L}_i , l \in \mathcal{C}_i
				\end{array}\right.\\
				&n_i^{j,l}=y_j^l+\sum_{h\in\mathcal{N}(j) , h \neq i}n_j^{h,l} , \ \forall l \in \mathcal{L} , j \in \mathcal{N}(i)\\
				&^{1}z_i^l={^{1}\bar{z}}_i^l=\max\{q_{\Phi(i)}^i,\hspace{-0.2cm}\max_{j \in \mathcal{G}_{k(i)}^l,j \neq i}y_j\} , \ \forall l \in \mathcal{L}_i , l \in \mathcal{C}_i\\
				&^{2}z_i^l={^{2}\bar{z}}_i^l=\textbf{1}_{\{q_{\Phi\left(i\right)}^i\}}\hspace{-0.1cm}\left(^{1}\bar{z}_i^l\right) +\hspace{-0.5cm}\sum_{j \in \mathcal{G}_{k\left(i\right)}^l,j \neq i}\hspace{-0.5cm}\textbf{1}_{\{y_j\}}\hspace{-0.1cm}\left(^{1}\bar{z}_i^l\right), \forall l \in \mathcal{L}_i , l \in \mathcal{C}_i\\
				& ^{2}a_i^{j,l}={^{1}a}_j^l , \ \forall j \in I_i, l \in \mathcal{L}_j
			\end{align}
		\end{subequations}
	\end{lemma}
	This lemma makes precise the comment made earlier about quadratic terms in the tax functions. At NE, agents force these quadratic terms to zero thus achieving consensus. Also, this lemma explains how summary messages are able to sum up the group demands of all agents using link $l$ at NE.
	\begin{lemma}(Primal Feasibility)\label{pfeas}
		At any $m \in \mathcal{NE}$ of the game $\mathfrak{G}$, the allocation vector $\hat{x}(m)$ is feasible.
	\end{lemma}

	\begin{lemma}\label{pw}
		At any $m \in \mathcal{NE}$ of the game $\mathfrak{G}$, the following constraints hold for all $i \in \mathcal{N}$ and $l \in \mathcal{L}_i$,
		\begin{subequations}
			\begin{align}
				&\hat{w}_i^l=\bar{w}_{-i}^l \label{eqw}\\
				&\bar{w}_{-i}^l(c^l-r_if_i^l)=0 \label{comw}\\
				&{^{2}p}_{\Phi(i)}^{i,l}({^{1}\bar{z}}_i^l-q_{\Phi(i)}^i)=0 \label{comp}
			\end{align}
		\end{subequations}
	\end{lemma}
	\vspace{0.3cm}
	
	Similarly, this Lemma shows how different groups form a consensus on the price of using each link $l$. Further, it provides the two complimentary slackness terms of the  KKT conditions~\eqref{eq:kkt}. In the proof of Theorem \ref{FI} we will see precisely how these expressions are utilized.
	\begin{lemma}\label{stationarity}
		The following constraints hold at any $m \in \mathcal{NE}$ of the game $\mathfrak{G}$,
		\begin{subequations}
			\begin{equation}
				v_i^\prime(\hat{x}_i(m)) = \sum_{l\in\mathcal{L}_i}{^{1}p}_i^l \quad \textrm{if}\quad \hat{x}_i(m)> 0,
			\end{equation}
			\begin{equation}
				v_i^\prime(\hat{x}_i(m)) \leq \sum_{l\in\mathcal{L}_i}{^{1}p}_i^l \quad \textrm{if}\quad \hat{x}_i(m)= 0.
			\end{equation}	
		\end{subequations}
		\label{stationarity}	
	\end{lemma}
	We will see how this lemma will be related to the stationarity term of the  KKT conditions~\eqref{eq:kkt} in the proof of Theorem \ref{FI}.
	
	\begin{lemma}(Individual Rationality and Weak Budget Balance)\label{IR}
		At any $m \in \mathcal{NE}$ of the game $\mathfrak{G}$, individual rationality is satisfied, $v_i(\hat{x}_i(m))-\hat{t}_i(m)\geq v_i(0), \ \forall i \in \mathcal{N}$.
		Also, the mechanism is weak budget balanced, $\sum_{i \in \mathcal{N}}\hat{t}_i(m)\geq 0$.
		
	\end{lemma}

	\begin{lemma}\label{exist}
		There exists a NE, $m \in \mathcal{NE}$, for the game $\mathfrak{G}$.
	\end{lemma}
	
	We are now ready to state the proof of Theorem~1.
	\begin{IEEEproof}[Proof of Theorem \ref{FI}]
		In the proof of Lemma \ref{exist}, we show that the message associated  to the solution of problem \eqref{CP2} is a NE of the game $\mathfrak{G}$. Now, we want to prove that all of the NE of the game $\mathfrak{G}$ generate allocation and prices that are efficient for problem \eqref{CP2}. Consider any $m\in \mathcal{NE}$, due to Lemmas \ref{quadratic}, \ref{pfeas}, \ref{pw} and \ref{stationarity}, the allocation vector, $\hat{x}(m)$, as $x^*$, $r_i{^{1}\bar{z}_i^l}$ as ${b_{k(i)}^{l^*}}$ (any $r_j{^{1}\bar{z}_j^l} ,j \in \mathcal{G}_{k(i)}^l$ could work too) and the variables ${^{1}p}_i^l$ and $w_i^l$ (or any $w_j^l$ for $j \in \mathcal{N}^l$) as ${\mu_i^l}^*$ and $\lambda_l^*$, respectively, satisfy the  KKT conditions~\eqref{eq:kkt}. Therefore, $\hat{x}(m)=x^*$ for any  $m\in \mathcal{NE}$ and hence, the allocation of all NE is unique and efficient. Also, due to Lemma \ref{exist}, we know at least one NE exists and therefore, the mechanism fully implements problem \eqref{CP2} or equivalently problem \eqref{CP} at its Nash equilibria.
		Furthermore, Lemma \ref{IR} proves individual rationality and weak budget balance properties.
	\end{IEEEproof}

	\section{Relaxing The Assumptions on Message Network} \label{sectionV}
	The primary reason of imposing Assumption \ref{assump} is that  there should be a consensus on the  prices of different groups using link $l$ at NE and this is not implementable by the proposed mechanism if the sub-graph of agents using link $l$ is not connected.  In this section, we propose an alternative extended mechanism that relaxes Assumption  \ref{assump}. The relaxed version of Assumption \ref{assump}, that is used in this section, is as follows.
	\begin{assumption}
		For each link $l \in \mathcal{L}$ and group $k \in \mathcal{K}^l$, there is at least one node $i \in \mathcal{G}_k^l$  that is connected to all other nodes $j \in \mathcal{G}_k^l$ and is denoted by $c(k,l)$.\label{assump2}
	\end{assumption}
	Note that Assumption \ref{assump2} only consists of the second part of Assumption \ref{assump}.
	
	In the alternative mechanism, we extend the agents that quote message $w_i^l$ from the agents using link $l$ to a bigger group of agents as follows.  For every link $l$, consider a connected sub-graph $\mathcal{GR}_l=(\mathcal{N}_l,\mathcal{E}_l)$ consisting of all agents $i \in \mathcal{N}^l$ in addition to the minimum number of other agents that do not use link $l$ and are required to make the sub-graph connected. This connected sub-graph is called link $l$'s sub-graph and we know that it exists due to the connectivity of the message graph.  For each agent $i$, the set of links $l \notin \mathcal{L}_i$ which $i \in \mathcal{N}_l$ are denoted by $\mathcal{L}^i$ with $|\mathcal{L}^i|=L^i$. The extended definition of message $w_i$ is $w_i=(w_i^l, l  \in \mathcal{L}_i\cup \mathcal{L}^i)$ and subsequently,  the definition of $\mathcal{N}^l(i)$ is modified as
	\begin{equation}
		\mathcal{N}^l(i)=
		\{j, j \in \mathcal{N}(i) \cap \mathcal{N}_l\} \ \forall i \in \mathcal{N}, l \in \mathcal{L}_i \cup \mathcal{L}^i.
	\end{equation}
	The tax functions are also modified. For $l \in \mathcal{L}^i$,
	\begin{equation}
		\begin{aligned}
			&\hat{t}_i^l(m)\hspace{-0.1cm}=\hspace{-0.2cm}\sum_{j\in\mathcal{N}(i)}(n_i^{j,l}-y_j^l-\hspace{-0.4cm}\sum_{h\in\mathcal{N}(j) , h \neq i} \hspace{-0.2cm}n_j^{h,l})^2
			\\
			&+(w_i^l-\bar{w}_{-i}^l)^2+\bar{w}_{-i}^l(w_i^l-\bar{w}_{-i}^l)(c^l-r_if_i^l)^2.
		\end{aligned}
	\end{equation}
	Since sub-graph of agents using each link $l$ may not be connected, we need other agents $i \notin \mathcal{N}^l$ to quote $w_i^l$ messages and help the agents $j \in \mathcal{N}^l$ in forming a consensus on the group prices of using link $l$. This is why two terms have been added to the tax function above that impose required conditions for the message $w_i^l$.
	The tax function does not change for $l \in \mathcal{L}_i$. For $l \notin \mathcal{L}_i \cup \mathcal{L}^i$, the tax function is the same as the $l \notin \mathcal{L}_i$ case for the original mechanism. It is straightforward to prove almost the same results  for this mechanism. Therefore, this mechanism also fully Nash implements problem \eqref{CP}, has individual rationality property at NE and is weak budget balanced at NE.

	\section{Conclusion} \label{section6}
	
	We proposed a distributed mechanism for the multicast transmission network that is applicable to a number of other scenarios with linear constraints and proved that it fully Nash implements the solution of problem \eqref{CP}. The main feature of this work is that message transmission is done locally via an underlying message network, in contrary to the standard mechanism design framework that allows message transmission throughout  the whole network.
	The dimensionality of agent $i$'s message in the main mechanism (with Assumption \ref{assump}) is $M_i=1+4L_i+N(i)L+|I_i|+2\sum_{j \in I_i}L_j+2|\mathcal{C}_i|$. Since the function $\Phi(i)$ chooses one agent $j \in \mathcal{N}(i)$, the average size of the set $I_i$ is $1$.  Also, we know that for each link $l$ and group $k \in \mathcal{K}^l$, there is one agent denoted by $c(k,l)$ and hence, the average size of $|\mathcal{C}_i|$ is $\frac{\sum_{l \in \mathcal{L}}K^l}{N}$.  Consequently, if we denote $\mathbb{E}_{i\in \mathcal{N}}(N(i))$ and $\mathbb{E}_{i\in \mathcal{N}}(L_i)$ by $\bar{N}$ and $\bar{L}$ respectively, the average size of the whole network's message is
	\begin{equation}
		\mathbb{E}(\sum_{i\in \mathcal{N}}M_i)\hspace{-0.1cm}=N(2+4\bar{L}+\bar{N}L+2\frac{\sum_{i \in \mathcal{N}}L_i}{N}+2\frac{\sum_{l \in \mathcal{L}}K^l}{N})
	\end{equation}
	which obviously grows linearly with the number of agents in the network, $N$.
	
	For the alternative mechanism, an extra term will be added to the message dimensionality of the whole network and that is $N\mathbb{E}_{i \in \mathcal{N}}(L^i)$. This is due to the extra messages quoted by agents to preserve the connectivity of message passing. 
	
	Message dimensionality in our mechanism is more efficient than the message dimensionality of the distributed mechanism proposed in \cite{sinha2017distributed} which grows with $N^2$ and it may be a consequence of learning guarantees that the proposed mechanism has.
	
	\bibliographystyle{IEEEtran}

	\section{Appendix}\label{appendix}
	
	\begin{IEEEproof}[Proof of Lemma \ref{con}]
		In order to show the strict concavity of $\hat{u}_i(m_i,m_{-i})$, we can show that its Hessian matrix, $\mathrm{H}$,  w.r.t. $m_i$ is negative definite and this is doable because the function $\hat{u}_i(m_i,m_{-i})$ is twice differentiable w.r.t $m_i$.
		It is obvious that cross derivatives of  $\hat{u}_i(m_i,m_{-i})$ w.r.t. different components of $m_i$ are zero which are the non-diagonal elements of $\mathrm{H}$. Hence, we consider the diagonal elements and show that they are all negative.

		It is straightforward to show that the second partial derivative of  $\hat{u}_i(m)$ w.r.t. all elements of messages $n_i$, $q_i$, $p_i$, $w_i$, $z_i$, $a_i$ and $\underline{y_i}$ is equal to $-2$. The only message element left is $y_i$ and the second partial derivative w.r.t. it is $\frac{\partial^2 \hat{u}_i(m)}{(\partial y_i)^2}=\frac{\partial^2 v_i(r_i{y_i})}{(\partial y_i)^2}=r_i\frac{\partial^2 v_i(\hat{x}_i)}{(\partial \hat{x}_i)^2}$. Since $v_i(\hat{x}_i)$ is strictly concave w.r.t. $\hat{x}_i$,  $\frac{\partial^2 \hat{u}_i(m)}{(\partial y_i)^2}<0$. Note that $r_i$ doesn't consist of any of agent $i$'s messages and so it is assumed as a constant.
		
		Therefore, since all of the diagonal elements of $\mathrm{H}$ are negative and non-diagonal elements are zero, matrix $\mathrm{H}$ is negative definite. We conclude that   $\hat{u}_i(m_i,m_{-i})$ is strictly concave w.r.t. $m_i$.
	\end{IEEEproof}

	\begin{IEEEproof}[Proof of Lemma \ref{quadratic}]
		At NE, every agent is best responding to other agents' messages and in order to find the relation between these messages, one can calculate the best response functions. Each of the results in this lemma corresponds to one of agent $i$'s messages and its relation with the messages of other agents. Therefore,  all of the results can be directly derived by setting each of their corresponding element of gradient to zero. For all agents $i \in \mathcal{N}$, we have
		\begin{equation*}
		\begin{aligned}
		\frac{\partial\hat{u}_i(m_i,m_{-i})}{\partial q_i^j}=0 \Rightarrow 2(q_i^j-y_j)=0
		\Rightarrow
		q_i^j=y_j,
		\forall j \in I_i
		\end{aligned}
		\end{equation*}
		As mentioned in the description of message components, $q_i^j$ can be used as a proxy for $y_j$ at NE and yet, agent $j$ can not change it.
		\begin{align*}
		&\frac{\partial\hat{u}_i(m_i,m_{-i})}{\partial(\underline{y_i}^l)}=0 \Rightarrow 2(\underline{y_i}^l-\frac{q_{\Phi(i)}^i\textbf{1}_{\{q_{\Phi(i)}^i\}}({^{1}\bar{z}}_i^l)}{{^{2}\bar{z}}_i^l})=0 \\
		&\Rightarrow
		\underline{y_i}^l=\frac{q_{\Phi(i)}^i\textbf{1}_{\{q_{\Phi(i)}^i\}}({^{1}\bar{z}}_i^l)}{{^{2}\bar{z}}_i^l},
		\ \forall  l \in \mathcal{L}_i \\
		&\frac{\partial\hat{u}_i(m_i,m_{-i})}{\partial (^{2}p_i^{j,l})}=0 \Rightarrow 2(^{2}p_i^{j,l}-{^{1}p}_j^l)=0
		\Rightarrow
		{^{2}p}_i^{j,l}={^{1}p}_j^l, \\
		&\hspace{6cm} \forall  j \in I_i, l \in \mathcal{L}_j\\
		&\frac{\partial\hat{u}_i(m_i,m_{-i})}{\partial w_i^l}=0 \Rightarrow 2(w_i^l-w_{c(k(i),l)}^l)=0 \\
		&\Rightarrow
		w_i^l=w_{c(k(i),l)}^l,
		\  \forall  l \in \mathcal{L}_i , l \notin \mathcal{C}_i\\
		&\frac{\partial\hat{u}_i(m_i,m_{-i})}{\partial w_i^l}=0 \Rightarrow 2(w_i^l-{^{2}p_{\Phi(i)}^{i,l}}-\hspace{-0.25cm}\sum_{j \in \mathcal{G}_{k(i)}^l, j \neq i}\hspace{-0.25cm}{^{1}p_j^l})=0\\
		&\Rightarrow
		w_i^l={^{2}p_{\Phi(i)}^{i,l}}+\sum_{j \in \mathcal{G}_{k(i)}^l, j \neq i}{^{1}p_j^l}, \
		\forall  l \in \mathcal{L}_i , l \in \mathcal{C}_i
		\end{align*}
		In all cases above, it is doable to have these relations because both right-hand sides of equations are non-negative.
		\begin{align*}
		&\frac{\partial\hat{u}_i(m_i,m_{-i})}{\partial n_i^{j,l}}=0 \Rightarrow 2(n_i^{j,l}-y_j^l-\sum_{h\in\mathcal{N}(j) , h \neq i}n_j^{h,l})=0 \\
		&\Rightarrow
		n_i^{j,l}=y_j^l+\sum_{h\in\mathcal{N}(j) , h \neq i}n_j^{h,l}, \ \forall  j \in \mathcal{N}(i), l \in \mathcal{L}
		\end{align*}
		Using a similar argument as the one used in \cite[p. 131]{sinha2017absi} we can prove that
		\begin{equation}
		n_i^{j,l}=\sum_{h\in\mathcal{N} , n(i,h)=j}y_h^l
		\end{equation}
		and consequently,
		\begin{equation}
		\sum_{j \in \mathcal{N}(i)}n_i^{j,l}=\sum_{h\in\mathcal{N},h \neq i}y_h^l
		\end{equation}
		Next, we show the remaining results that are about the message elements of $z_i$ and ${^2a}_i$.
		
		\begin{align*}
		&\frac{\partial\hat{u}_i(m_i,m_{-i})}{\partial ({^{1}z}_i^l)}=0 \Rightarrow 2({^{1}z}_i^l-{^{1}\bar{z}}_i^l)=0
		\Rightarrow
		{^{1}z}_i^l={^{1}\bar{z}}_i^l\\
		&\hspace{5.8cm} \forall  l \in \mathcal{L}_i , l \in \mathcal{C}_i \nonumber\\
		&\frac{\partial\hat{u}_i(m_i,m_{-i})}{\partial ({^{2}z}_i^l)}=0 \Rightarrow 2({^{2}z}_i^l-{^{2}\bar{z}}_i^l)=0
		\Rightarrow
		{^{2}z}_i^l={^{2}\bar{z}}_i^l \\
		&\hspace{5.8cm}  \forall  l \in \mathcal{L}_i , l \in \mathcal{C}_i \nonumber\\
		&\frac{\partial\hat{u}_i(m_i,m_{-i})}{\partial ({^{2}a}_i^{j,l})}=0  \Rightarrow \hspace{-0.05cm} 2(^{2}a_i^{j,l}-\hspace{-0.1cm}{^{1}a}_j^l)\hspace{-0.05cm}=\hspace{-0.05cm}0
		\Rightarrow
		{^{2}a}_i^{j,l}\hspace{-0.1cm}=\hspace{-0.1cm}{^{1}a}_j^l \\
		&\hspace{5.8cm} \forall  j \in I_i, l \in \mathcal{L}_j \nonumber
		\end{align*}
	\end{IEEEproof}
	
	\begin{IEEEproof}[Proof of Lemma \ref{pfeas}]
		According to Lemma \ref{quadratic}, the following relation holds at NE,
		\begin{equation*}
		f_i^l=\sum_{j \in \mathcal{N}}y_j^l
		\end{equation*}
		Note that it implies that all of the agents $i \in \mathcal{N}$ have the same $f_i^l$ and consequently the same $r_i$ at NE. Further, due to Lemma \ref{quadratic},
		\begin{equation*}
		y_j^l=\left\{
		\begin{array}{cc}
		\frac{y_j}{n_{\textrm{max}}^{k,l}} & \ \textrm{if} \ l \in \mathcal{L}_j, \ y_j=\max_{i \in \mathcal{G}_{k(j)}^l}\{y_i\}\\
		0 & \ \textrm{oth.}
		\end{array}\right.
		\end{equation*}
		where $n_{\textrm{max}}^{k,l}$ is the number of agents $j \in \mathcal{G}_k^l$ with $y_j=\max_{i \in \mathcal{G}_{k(j)}^l}\{y_i\}$.
		Consequently, $\forall l \in \mathcal{L}$, we can write
		\begin{equation*}
		\begin{aligned}
		&\sum_{k\in \mathcal{K}^l} \max_{i \in \mathcal{G}_k^l} \{\hat{x}_i\}=\hspace{-0.2cm}\sum_{k\in \mathcal{K}^l} \max_{i \in \mathcal{G}_k^l}\{r_iy_i\} \hspace{-0.1cm}  &\\ & \leq \frac{c^l}{\sum_{j \in \mathcal{N}}y_j^l}\sum_{k\in \mathcal{K}^l} \max_{i \in \mathcal{G}_k^l}\{y_i\}
		=\frac{c^l}{\sum_{j \in \mathcal{N}}y_j^l}\sum_{i \in \mathcal{N}}y_i^l=c^l
		\end{aligned}
		\end{equation*}
		Which proves that the allocation at NE is feasible.	
	\end{IEEEproof}
	
	\begin{IEEEproof}[Proof of Lemma \ref{pw}]
	We first prove result \eqref{eqw}. 
		Assume $\exists i \in \mathcal{N}, l \in \mathcal{L}_i$ so that $\hat{w}_i^l \neq \bar{w}_{-i}^l$. Since, $w_i^l=\hat{w}_i^l$ at NE, it implies that $ \exists i,j \in \mathcal{N}^l, \
		\hat{w}_i^l \neq \hat{w}_j^l$ and therefore there exists an agent $h \in \mathcal{N}^l$ for which $\hat{w}_h^l>\bar{w}_{-h}^l$ or equivalently, $\hat{w}_h^l=\bar{w}_{-h}^l+\epsilon$ for some $\epsilon>0$. We will show that agent $h$ has a profitable deviation by decreasing his message ${^1a}_h^l$ to ${^{1}a}_h^{l'}={^{1}a}_h^l-\epsilon'>0$ for some $0<\epsilon'<\epsilon$. Consequently ${\hat{w}_h^{l'}}=\hat{w}_h^l-\epsilon'=\bar{w}_{-h}^l+\epsilon-\epsilon'=\bar{w}_{-h}^l+\epsilon''$. We can write
		\begin{align*}
		&\hat{u}_h(.,{^{1}a}_h^{l'})
		-\hat{u}_h(.,{^{1}a}_h^l)=\\&
		-\epsilon''^2-\bar{w}_{-h}^l\epsilon''(c^l-r_hf_h^l)^2+\epsilon^2+\bar{w}_{-h}^l\epsilon(c^l-r_hf_h^l)^2\\&=\underbrace{\epsilon^2-\epsilon''^2}_{>0}+
		\underbrace{\bar{w}_{-h}^l(\epsilon-\epsilon'')(c^l-r_hf_h^l)^2}_{\geq 0}>0
		\end{align*}
		and we conclude that at any NE, $\hat{w}_i^l=\bar{w}_{-i}^l, \ \forall i \in \mathcal{N} , l \in \mathcal{L}_i$.
		
		Now we prove result \eqref{comw}. Suppose $\exists i \in \mathcal{N}, l \in \mathcal{L}_i$ so that $\bar{w}_{-i}^l(c^l-r_if_i^l) \neq 0$. It implies $\bar{w}_{-i}^l>0$ and $(c^l-r_if_i^l)^2>0$. We show that agent $i$ benefits from deviating to ${^{1}a}_i^{l'}={^{1}a}_i^l-\epsilon>0$, for some $\epsilon>0$. According to the first result of this lemma, $\bar{w}_i^l=\bar{w}_{-i}^l$ and we have
		\begin{align*}
		&\hat{u}_i(.,{^{1}a}_i^{l'})
		-\hat{u}_i(.,{^{1}a}_i^l)=
		-\epsilon^2+\bar{w}_{-i}^l\epsilon(c^l-r_if_i^l)^2
		\\&=\epsilon(-\epsilon+\underbrace{\bar{w}_{-i}^l(c^l-r_if_i^l)^2}_{> 0,\  \textrm{Due to assumption}})=\epsilon(-\epsilon+\alpha)>0, \  \textrm{for} \ \epsilon<\alpha
		\end{align*}
		Since $\alpha>0$ , profitable deviation by positive $\epsilon$ is possible and the result is proved.
		
		Proving the result \eqref{comp} is similar to the result \eqref{comw}. Assume $\exists i \in \mathcal{N}, l \in \mathcal{L}_i$ so that ${^{2}p}_{\Phi(i)}^{i,l}({^{1}\bar{z}}_i^l-q_{\Phi(i)}^i) \neq 0$. It implies that ${^{2}p}_{\Phi(i)}^{i,l}>0$ and $({^{1}\bar{z}}_i^l-q_{\Phi(i)}^i)^2>0$. Due to Lemma \ref{quadratic},  ${^{2}p}_{\Phi(i)}^{i,l}={^{1}p}_i^l$ and so ${^{1}p}_i^l>0$. We prove agent $i$ has a profitable deviation to  ${^{1}p}_i^{l'}\hspace{-0.1cm}=\hspace{-0.05cm}{^{1}p}_i^l-\epsilon\hspace{-0.05cm}>0$, for some $\epsilon>0$.
		\begin{align*}
		&\hat{u}_i(.,{^{1}p}_i^{l'})
		-\hat{u}_i(.,{^{1}p}_i^l)=
		-(\epsilon)^2+{^{2}p}_{\Phi(i)}^{i,l}\epsilon(^{1}\bar{z}_i^l-q_{\Phi(i)}^i)^2\\
		&\hspace{-0.1cm}=\epsilon(-\epsilon+\underbrace{{^{2}p}_{\Phi(i)}^{i,l}(^{1}\bar{z}_i^l-q_{\Phi(i)}^i)^2}_{> 0, \ \textrm{Due to assumption}})\hspace{-0.05cm}=\epsilon(-\epsilon+\alpha)\hspace{-0.05cm}>\hspace{-0.05cm}0,  \textrm{for} \ \epsilon<\alpha
		\end{align*}
		Since $\alpha>0$, agent $i$ can profit by deviating with positive $\epsilon$ and the result is proved.
	\end{IEEEproof}
	
	\begin{IEEEproof}[Proof of Lemma \ref{stationarity}]
		If $\hat{x}_i(m)>0$, then $y_i>0$ and hence, the partial derivative of $\hat{u}_i(m_i,m_{-i})$ w.r.t. $y_i$ must be zero at NE. Therefore,
		\begin{equation*}
		\begin{aligned}
		&\frac{\partial\hat{u}_i(m_i,m_{-i})}{\partial y_i}=0
		\Rightarrow (\frac{\partial\hat{u}_i(m_i,m_{-i})}{\partial \hat{x}_i(m)})\frac{d\hat{x}_i(m)}{dy_i}=0 \\
		&\Rightarrow\hspace{-0.05cm}
		(v_i^\prime(\hat{x}_i(m))-\hspace{-0.05cm}\sum_{l\in\mathcal{L}_i}{^{2}p}_{\Phi(i)}^{i,l})r_i=0
		\Rightarrow\hspace{-0.05cm} v_i^\prime(\hat{x}_i(m))=\hspace{-0.05cm}\sum_{l\in\mathcal{L}_i}{^{1}p}_i^l
		\end{aligned}
		\end{equation*}
		Note that $r_i\neq0$.
		If $\hat{x}_i=0$, then $y_i=0$ and therefore, the partial derivative of $\hat{u}_i(m_i,m_{-i})$ w.r.t. $y_i$ must not be positive at NE. Hence,
		\begin{equation*}
		\begin{aligned}
		&\frac{\partial\hat{u}_i(m_i,m_{-i})}{\partial y_i}\leq0
		\Rightarrow (\frac{\partial\hat{u}_i(m_i,m_{-i})}{\partial \hat{x}_i(m)})\frac{d\hat{x}_i(m)}{dy_i}\leq0 \\
		&\Rightarrow \hspace{-0.05cm}
		(v_i^\prime(\hat{x}_i(m))-\hspace{-0.05cm}\sum_{l\in\mathcal{L}_i}{^{2}p}_{\Phi(i)}^{i,l})r_i\leq0
		\Rightarrow v_i^\prime(\hat{x}_i(m))\hspace{-0.05cm}\leq\hspace{-0.05cm}\sum_{l\in\mathcal{L}_i}{^{1}p}_i^l
		\end{aligned}
		\end{equation*}
	\end{IEEEproof}
	\begin{IEEEproof}[Proof of Lemma \ref{IR}]
		At NE, we can write $\hat{t}_i=\hat{x}_i(m)\sum_{l \in \mathcal{L}_i}{^{1}p}_i^l$ and hence, $\sum_{i \in \mathcal{N}}\hat{t}_i\geq 0$ and mechanism is weak budget balanced.
		
		The individual rationality is obvious for $\hat{x}_i(m)=0$. For  $\hat{x}_i(m)>0$, the function $u_i$ is defined as follows,
		\begin{equation*}
		u_i(x)=v_i(x)-x\sum_{l \in \mathcal{L}_i}{^{1}p}_i^l
		\end{equation*}
		Since $u_i(x)$ is strictly concave w.r.t. $x$ and $u_i^{\prime}(\hat{x}_i(m))=0$, $u_i^{\prime}(y)>0$ for $0\leq y<\hat{x}_i(m)$. We can conclude $u_i(y)\geq u_i(0)$ and due to continuity, $u_i(\hat{x}_i(m))\geq u_i(0)$. Since $u_i(0)=v_i(0)$ and $u_i(\hat{x}_i(m))=v_i(\hat{x}_i(m))-\hat{t}_i(m)$, we conclude that 	$v_i(\hat{x}_i(m))-\hat{t}_i(m)\geq v_i(0)$ and the result is proved.
	\end{IEEEproof}

			\begin{IEEEproof}[Proof of Lemma \ref{exist}]
				To prove the existence of a NE, we prove that a suggested valid message is a NE.  This suggested message is generated based on the solution of problem \eqref{CP2} which we know exists and is unique.
				We notice that because of monotonically increasing property of valuation functions, the solution of problem \eqref{CP2} always lies in the Pareto optimal region of feasible set which in our case, is the upper boundary of feasible set. Suppose $(x^*,b^*,\lambda^*,\mu^*)$ is the solution of problem \eqref{CP2}. We generate $m$ as following. First assume $m$ satisfies all of the constraints in Lemma \ref{quadratic}. Further, $y$ is set to be any scaled version of $x^*$ and since $x^*$ is on the boundary of feasible region, $\hat{x}(m)=x^*$. In addition, $^{1}p_i^l$ is set to be equal to ${\mu_i^l}^*$ and this is valid since ${\mu_i^l}^*\geq0$. Then, $w_i^l=\sum_{j \in \mathcal{G}_{k(i)}^l}{^{1}p_j^l}=\sum_{j \in \mathcal{G}_{k(i)}^l}{\mu_j^l}^*=\lambda_l^*$. Also $r_i{^{1}\bar{z}}_i^l=\max_{j \in \mathcal{G}_{k(i)}^l}\{r_iy_j\}=\max_{j \in \mathcal{G}_{k(i)}^l}\{\hat{x}_i\}=\max_{j \in \mathcal{G}_{k(i)}^l}\{x_i^*\}={b_k^l}^*$. Hence, Lemma \ref{pw} is satisfied for $m$. Also, due to stationarity condition, Lemma \ref{stationarity} is satisfied for $m$. In sum, since Lemmas  \ref{quadratic}, \ref{pw} and \ref{stationarity} are satisfied, we know that the elements of the gradient vector of utility function of each agent $i$ w.r.t. $m_i$ is either zero (positive messages) or not positive (zero messages) which implies that each agent is best responding to other agents' messages and therefore, $m$ is a NE.
			\end{IEEEproof}
\end{document}